\documentclass[conference]{IEEEtran}
\usepackage{resizegather}
\usepackage{url}
\usepackage{cite}
\usepackage{amsmath}
\usepackage{graphicx}
\usepackage{multirow}
\usepackage{blindtext}
\usepackage{subcaption}
\usepackage{amssymb,mathtools,bm}
\usepackage[margin=0.605in,top=0.7in,bottom=0.935in]{geometry}
\usepackage{algorithm,algorithmic}
\usepackage[english]{babel}
\usepackage[table,xcdraw]{xcolor}
\usepackage[hidelinks,bookmarks=false]{hyperref}
\usepackage{booktabs}

\usepackage{subcaption}
\usepackage{xcolor}

\newcommand{\comment}[1]{}

\renewcommand{\arraystretch}{0.8}
\makeatletter
\renewcommand{\fnum@figure}{Fig. \thefigure}
\makeatother

\begin{document} 
\title{Positioning Error Impact Compensation through Data-Driven Optimization in User-Centric Networks}
\author{\IEEEauthorblockN{Waseem Raza\IEEEauthorrefmark{1}, Fahd Ahmed Khan\IEEEauthorrefmark{1}, Muhammad Umar Bin Farooq\IEEEauthorrefmark{1}, Sabit Ekin\IEEEauthorrefmark{2} and Ali Imran\IEEEauthorrefmark{1}\IEEEauthorrefmark{3}}
\IEEEauthorblockA{\IEEEauthorrefmark{1}AI4Networks Research Center, School of Electrical \& Computer Engineering, University of Oklahoma, OK, USA\\
\IEEEauthorrefmark{3}{James Watt School of Engineering, University of Glasgow, United Kingdom.}
}
\IEEEauthorblockA{\IEEEauthorrefmark{2}School of Electrical and Computer Engineering, {Texas A\&M University}, TX, USA.\\
Email: \{waseem, fahd.khan, umar.farooq, ali.imran\}@ou.edu, {sabitekin@tamu.edu}}}
\maketitle
 
\begin{abstract}
The performance of user-centric ultra-dense networks (UCUDNs) hinges on the Service zone (Szone) radius, which is an elastic parameter that balances the area spectral efficiency (ASE) and energy efficiency (EE) of the network. Accurately determining the Szone radius requires the precise location of the user equipment (UE) and data base stations (DBSs). Even a slight error in reported positions of DBSs or UE will lead to an incorrect determination of Szone radius and UE-DBS pairing, leading to degradation of the UE-DBS communication link. To compensate for the positioning error impact and improve the ASE and EE of the UCUDN, this work proposes a data-driven optimization and error compensation (DD-OEC) framework. The framework comprises an additional machine learning model that assesses the impact of residual errors and regulates the erroneous data-driven optimization to output Szone radius, transmit power, and DBS density values which improve network ASE and EE. The performance of the framework is compared to a baseline scheme, which does not employ the residual, and results demonstrate that the DD-OEC framework outperforms the baseline, achieving up to a $23\%$ improvement in performance.

\comment{
The performance of the user-centric ultra-dense networks (UCUDNs) is degraded due to erroneous position estimates of user equipment (UE) and data base station (DBS). These errors infiltrate the databases utilized by the CC for data-driven network operation and optimization. Erroneous databases can result in sub-optimal configuration and optimization parameters (COPs) in data-driven optimization. This paper proposes a framework to adopt the data-driven approach to solve the issue of sub-optimal COPs. To realize this, the framework utilizes residual error modeling in the optimization engine (OE). For this purpose, an error-free historical database is maintained, and the automated machine learning module uses the residual data. The multi-objective optimization problem is solved using two meta-heuristic techniques, the genetic algorithm and simulated annealing for various optimization scenarios. Results, shown in terms of the value and required iterations to reach the objective function convergence, validate the efficacy of the proposed framework in mitigating the impact of positioning error on optimization problems.}
\end{abstract}
\begin{IEEEkeywords}
User-centric ultra-dense networks (UCUDNs), Positioning errors, Machine Learning, Residual learning, and multi-objective optimization.
\end{IEEEkeywords}
\IEEEpeerreviewmaketitle
\section{Introduction}
Emerging wireless networks are adopting dense deployment to support an array of applications with diverse requirements \cite{BibDensification}. The cellular network is required to simultaneously facilitate high data-rate extended reality applications, ultra-reliable low-latency communication for intelligent transportation systems, and energy-constrained devices with low data rates, such as surveillance systems or body area networks \cite{Bib5Gapplications}. The dense deployment of base stations and heterogeneity of services make network management increasingly complex as the number of configuration parameters grows exponentially. In this context, Artificial Intelligence (AI) and Machine Learning (ML) techniques demonstrate breakthrough performances, leading to an era of AI-based wireless network configuration as an inevitable future~\cite{R06_18_DBCM5G_TBD,R07_19_WNDDLMAI_TCOM}. Although mathematical and analytical techniques are still useful, data-driven models undoubtedly play a complementary role in improving future network design and operational performance. Hence, data-driven modeling is employed for mobility prediction, proactive handovers, network fault diagnosis, and other tasks~\cite{R08_21_RADDWC_COMST}.

In addition to incorporating intelligence into emerging networks, new architectures for 5G and beyond networks that can quickly adapt to changing service requirements are being explored. One such elastic architecture is the user-centric architecture with flexible service zones, which is gaining recognition due to its promising benefits, such as lower interference, improved area spectral efficiency, and better quality of experience (QoE) by eliminating cell-edge users~\cite{R04_18_UCRAN_Access, BibShahrukhFlexibleSzonePaper}. These benefits are realized by serving users within non-overlapping flexible service zones (Szones) based on their service requirements. The Szone of a user-centric ultra-dense network (UCUDN) and serving data base station (DBS) activation is controlled by a central controller, which requires accurate user equipment (UE) and DBS location information for efficient network operation and resource allocation.

While accurate location information is also necessary for base station (BS) centric networks in tasks involving network automation and self-healing, its importance increases in UCUDNs. Even a slight error in the position of the UE or the DBS can result in scheduling UEs that are in close proximity to each other or activating DBSs that are far from the UEs. Both scenarios can result in high interference, which can have a detrimental effect on network performance. Therefore, in real-world UCUDNs, where the central controller may not have access to error-free locations of all UEs and DBS, the use of AI and ML techniques to compensate for imperfect location information is essential for its data-driven operations~\cite{R09_19_HSCAUC_WCSP,R10_20_MLPP_LWC}.

Characterization of the impact of positioning errors in traditional BS-centric networks has been carried out in several studies~\cite{R12_16_MDTACE_WCL,R13_16_PEASE_ICC, R14_19_OBWACE_COML}. The reliability of the data-based autonomous coverage estimation in the presence of UE and BS positioning error was investigated and quantified in \cite{R12_16_MDTACE_WCL}. Onireti et. al.,~\cite{R13_16_PEASE_ICC} demonstrated that inaccuracies in positioning techniques could result in faulty associations, leading to degraded area spectral efficiency (ASE). This issue is further exacerbated by the increased BS density in ultra-dense networks. In~\cite{R14_19_OBWACE_COML}, the authors demonstrated that an optimal bin width could mitigate the effects of positioning and quantization errors, resulting in the most accurate coverage estimate.

Most existing works on UCUDNs assume accurate location information for UEs and DBS \cite{R10_20_MLPP_LWC,R05_20_EDQEE_GCN,R09_19_HSCAUC_WCSP}, and there has been very little work on investigating and understanding the impact of positioning error. Erroneous position estimates were considered in \cite{Waseem-WCNC-2022}, where time-series forecasting was utilized to predict the trend of received signal power, ASE, and energy efficiency (EE) values, with varying three configuration and optimization parameters (COPs), namely, transmit power, Szone radius, and DBS density. The prediction was restricted within a prediction window, and only a limited number of COP combinations were explored. Moreover, the error prediction was not utilized to determine the optimal COP combination values, which jointly optimize the network performance. In this work, we adopt a different approach compared to \cite{Waseem-WCNC-2022}, and an AutoML model is trained to learn the impact of position error on network key performance indicators (KPIs). This error characterization is then utilized to compensate for the joint KPI optimization and yield improved network configuration values of transmit power, Szone radius, and DBS density. The contribution of this paper is summarized below. 
\begin{itemize}
    \item We propose a data-driven optimization and error compensation (DD-OEC) framework to jointly maximize two KPIs, namely, ASE and EE, where UE service is impacted by the positioning error. The framework adopts a data-driven approach and utilizes an AutoML module for learning the relationship between the COPs and the KPIs. The trained ML model is then employed for heuristic multi-objective optimization.  
    \item We demonstrate that the candidate COP solution derived from data with positioning errors is suboptimal. To mitigate this adverse impact of positioning errors, we employ a novel AutoML model to learn the \emph{residual} error obtained from the difference between the current erroneous database and a historical error-free database. This residual error is then utilized during the optimization process to compute COP values that improve the corresponding KPI values.
    \item The performance of the proposed DD-OEC is compared with a baseline scheme that does not take into account the \emph{residual}. Simulation results reveal that the proposed DD-OEC scheme is able to learn the residual and incorporate it to perform better than the baseline scheme. Therefore, the DD-OEC optimization is necessary to compensate for the error impact on the optimization solution.
\end{itemize}

\begin{figure*}[t!]
	\centering
	\begin{subfigure}{0.23\textwidth}
		\centering
		\includegraphics[width= 4.950cm, height=5cm]{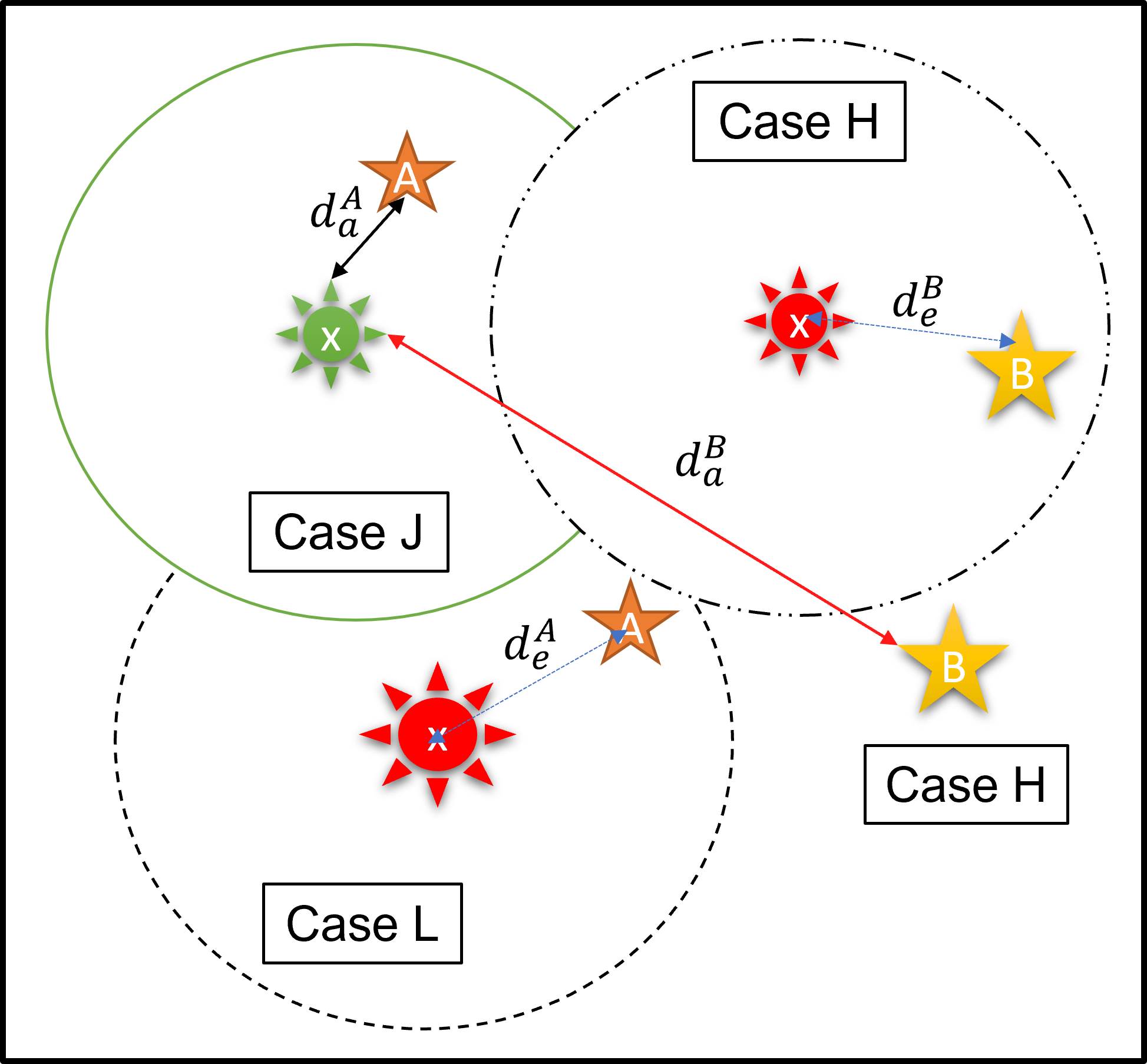}
		\caption{Positioning error modeling.}
		\label{fig:pos-error_model}
	\end{subfigure}
	\begin{subfigure}{0.76\textwidth}
		\centering
		\includegraphics[width=12.750cm, height=5cm]{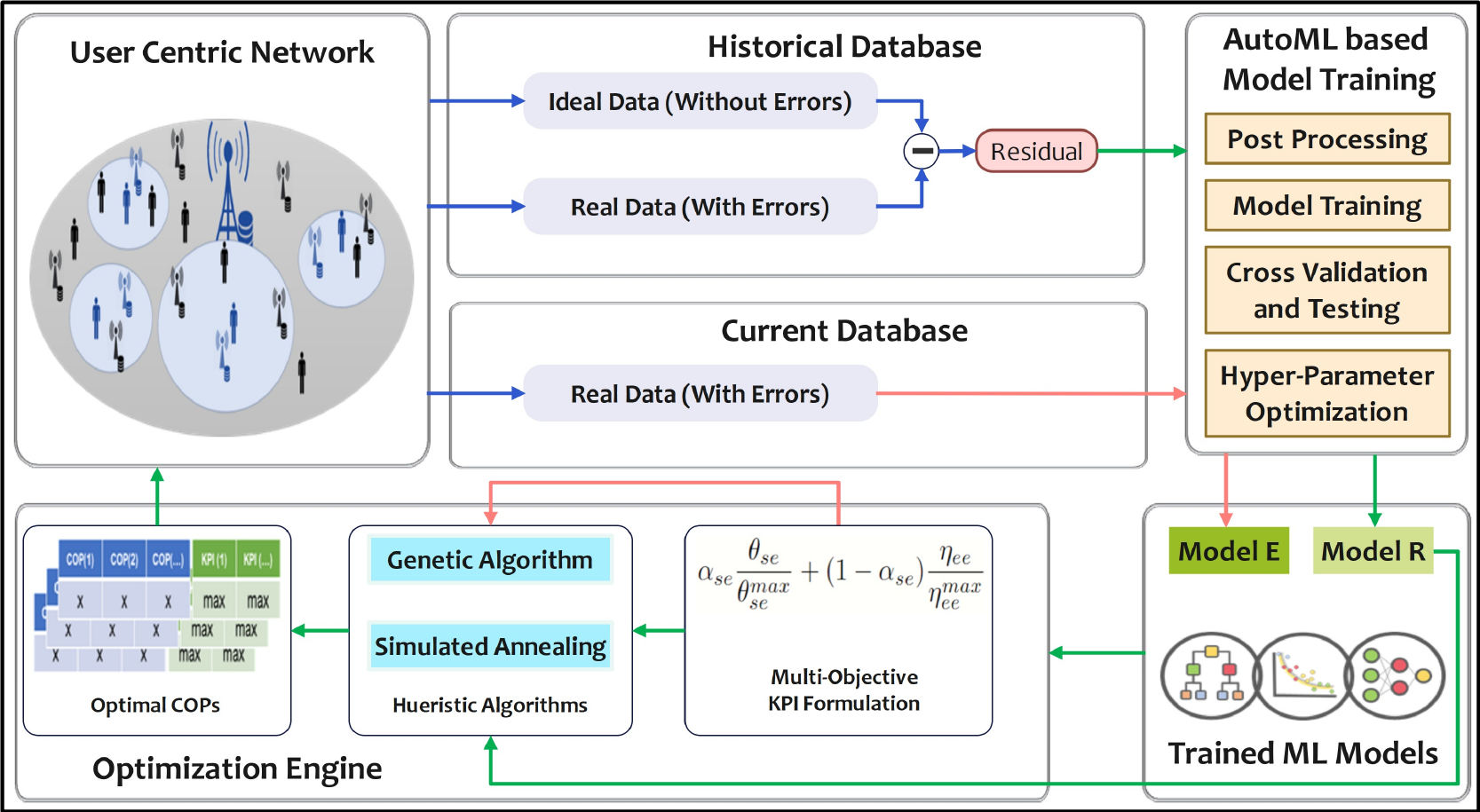}
		\caption{The \underline{d}ata-\underline{d}riven \underline{o}ptimization and \underline{e}rror \underline{c}ompensation modeling (DD-OEC) framework.}
		\label{fig:system-model-framework}
	\end{subfigure}
	\caption{{Positioning error modeling, and the DD-OEC framework for the data-driven optimization and error compensation COP-KPI optimization. This framework includes synthetic data generation, AutoML based model fitting, and multi objective optimization.}}
	\label{fig:error-model-framework}
 \vspace{-0.65cm}
\end{figure*}
\section{System Model}\label{sec:system-model-prelim}
We consider a UCUDN, where the DBSs and the UEs are assumed to be distributed randomly over an area using a Poisson point process. The density of the DBSs and the UEs is represented by $\lambda_{dbs}$ and  $\lambda_{ue}$, respectively. All the DBS are equipped with a single omnidirectional antenna and transmit with the same power. A macro BS acts as the central controller, collecting data from the UEs and scheduling them during each transmission time interval. The controller creates virtual non-overlapping Szones, of radius $R_{sz}$, around the scheduled UEs and activates a DBS within each Szone to serve the respective UEs as shown in Fig.~1b. This ensures that there is at least a distance greater than $2R_{sz}$ between the scheduled UEs. Moreover, the UE is served by the DBS, having the best channel power, located within a distance of $R_{sz}$ from the UE. Consequently, if the distance between UEs is less than $2 R_{sz}$, then they are scheduled in different TTIs based on scheduling priority. This scheduling and activation criterion limits the interference to the serviced UEs. Moreover, if a UE has no DBS within its Szone, the network can adjust and increase $R_{sz}$ to serve the disadvantaged UEs.
\subsection{Channel Model}
We adopt the close-in two-slope path-loss model \cite{path_loss_shad},
\small
\begin{equation}
\label{eq:PL}
PL(d) = -F(f,1m)_{dB} - 10l_1\log d - 10l_2\log\left(\frac{d}{d_{t}}\right)u( d - d_{t}),
\end{equation}\normalsize
where $d$ is the 3D distance between the UE and BS, $F(f,1m)_{[dB]}=20\log_{10}\big(\frac{4\pi f}{c}\big)$ is the free space pathloss in dB at a transmitter-receiver distance of 1 meter at carrier frequency $f$ with $c$ representing the speed of light, $u(\cdot)$ denotes the unit step function, $d_{t}$ is the breakpoint or threshold distance, $l_1$ and $l_2$ are the path loss exponents for a distance less than $d_{t}$ and greater than $d_{t}$, respectively. The signal-to-interference plus noise ratio (SINR) for UE $x$ is given as,
\begin{equation}
\gamma_x = \frac{P_x G_x \chi_x 10^{\frac{{PL}(d_x)}{10}}}{N_0 + \sum\limits_{\forall i\in \mathbb I} P_i G_i \chi_i 10^{\frac{{PL}(d_i)}{10}}},
\end{equation}
where $P_x$ is the transmit power of the serving DBS, $G_x$ is the downlink antenna gain of the serving DBS, $\chi_x$ denotes the channel shadowing modeled as a log-normal distribution with $0$dB mean and $4$dB variance, $d_x$ is the distance between the serving DBS and UE $x$. $N_0$ is the thermal noise power, $P_i$ and $G_i$ are the transmit power and antenna gain of the $i$-th interfering DBS, respectively. $\chi_i$ shows the channel shadowing for interfering BS $i$ modeled as a log-normal distribution with $0$dB mean and $4$dB variance, and $d_i$ is the distance between UE $x$ and the $i$-th interfering DBS. $\mathbb I$ is the set of all interfering BS.
\subsection{Key Performance Indicators:}

\par The performance of the UCUDN is determined based on area spectral efficiency and network energy efficiency. ASE is defined as the number of bits that are transmitted to a UE from a DBS per unit bandwidth per unit area and is mathematically expressed as $\theta_{se} = \frac{1}{A}\sum\limits_{\forall x\in \mathbb U}\log_{2}(1+\gamma_x)$, where $\mathbb U$ is the set of all served UEs in the network and $A$ is the simulation area. EE is defined as $\eta_{ee} = \frac{A}{P_T} \theta_{se}$, where $P_T$ is the total power consumption of all CBS and DBS in the network following the network energy consumption model in~\cite{Earth_project}.

\subsection{Positioning Error Impact and its Modeling}\label{sec:pei_and_mod}
\par In UCUDN, the central controller requires exact location information of the UEs and DBSs for scheduling, DBS activation, and determining $R_{sz}$. However, in practical systems, only estimates of the positions are available\footnote{Although there can be various other sources of imperfections such as channel modeling errors, quantization errors, etc. however, the scope of this paper is limited to the impact of error in UE and DBS positions.}. These positioning errors affect UE scheduling, DBS activation, and Szone radius. For instance, it can lead to the scheduling of UEs that are spaced less than $2 R_{sz}$ apart, thus, increasing the interference and lowering the QoE. Similarly, a DBS can be activated which is outside the Szone of the UE, again resulting in lower reference signal received power (RSRP) and increased interference to other UEs.
\comment{
Apart from the fall in user QoE, positioning errors can have adverse effects on data-driven optimization results, which will be discussed in Section IV.
}

For error modeling, we assume that the ideal position of a DBS or UE will be uniformly distributed within the circular ring of error-radius, $R_{er}$, centered around the position estimates. A lower value of $R_{er}$ will indicate a more accurate estimate and vice versa. In the subsequent discussion, we define \emph{perceived distance} as the distance between the UE and DBS based on the position estimates and \emph{actual distance} as the true physical distance. The former is shown by the subscript `e', and the latter by `a' in the distance \textit{d} in the following discussion. 

To elaborate the positioning error impact, we assume a UE $x$ and two DBSs $A$ and $B$ as shown in Fig.~\ref{fig:pos-error_model}. In the ideal case, referred as \textit{Case J}, the actual and reported positions of UE $x$ and DBS $A$ are the same; hence the Szone is shown by a green solid circle in Fig.~\ref{fig:pos-error_model}, and the SINR depends on the actual distance $d_{a}^{A}$. However, with positioning errors, two scenarios occurs based on the error magnitude, as shown in Fig.~\ref{fig:pos-error_model}. 
\begin{itemize}
\item In low error regimes, shown as \emph{Case L}, despite the position errors, the UE-DBS association is not changed and remains the same as the ideal case. However, despite this ideal-like association, the central controller allocates resources based on the perceived distance, $d_{e}^{A}$, which is different from the actual distance $d_{a}^{A}$. 

\item In the high error regimes, the positioning errors cause a change in UE-DBS association, i.e., UE $x$ is now associated with the distant DBS $B$ instead of DBS $A$ due to lower perceived distance $d_e^{B}$, as shown in \emph{Case H} in the figure. It is possible that in this case, the DBS $B$ might actually lie outside of the S-zone. As a result, in addition to the sub-optimal resource allocation (based on the perceived distance), the larger value of $d_a^{B}$, leads to a significant deterioration of SINR at the UE.  
\end{itemize}
Thus, from this discussion, it can be summarized that positioning errors can result in significant variations in the perceived distances between UEs and DBSs. This, in turn, can lead to suboptimal resource allocation and UE-DBS association, negatively affecting the network's KPIs of interest.

\section{Optimization Problem Formulation and DD-OEC Framework}
In this Section, we present the optimization objective function and introduce the proposed framework for data-driven optimization and error compensation (DD-OEC).

\subsection{Optimization Problem Formulation and Algorithm}
\par As mentioned earlier, our focus in this study is to enhance both the ASE and EE. Since these two KPIs exhibit a certain degree of trade-off, we develop a multi-objective optimization function for the data-driven optimization process. The objective function is formulated as the weighted sum of the normalized KPIs, which is specifically defined as,
\begin{equation}\label{eq:obj_func_Eq}
f_{obj}(\theta_{se},~\eta_{ee}) = \alpha_{se}\frac{\theta_{se}}{\theta_{se}^{max}}+(1-\alpha_{se})\frac{\eta_{ee}}{\eta_{ee}^{max}},
\end{equation}
where $\alpha_{se}$ and $1-\alpha_{se}$ are the weight factors for ASE and EE, respectively. These KPIs are functions of the COPs; DBS density, Szone radius, and transmit power. Since maximizing both ASE and EE is our goal, the joint optimization problem can be expressed as follows,
\begin{equation}\label{eq:moop_formul}
	\begin{aligned}
		\max_{\lambda_{dbs},~R_{sz},~P_{tx}} \quad & f_{obj}(\theta_{se},~\eta_{ee}),\\
		\textrm{s.t.} \quad & 0\leq\lambda_{dbs}^{min}\leq\lambda_{dbs}\leq\lambda_{dbs}^{max},\\
		& 0\leq R_{sz}^{min}\leq R_{sz}\leq R_{sz}^{max},\\    
		& 0\leq P_{tx}^{min}\leq P_{tx}\leq P_{tx}^{max}.\\    		                    
	\end{aligned}
\end{equation}
where, $\lambda_{dbs}^{min}$ and $\lambda_{dbs}^{max}$ denote the minimum and maximum value of DBS density, $R_{sz}^{min}$ and $R_{sz}^{max}$ denote the minimum and maximum value of Szone radius, and $R_{sz}^{min}$ and $R_{sz}^{max}$ denote the minimum and maximum value of Szone radius, respectively. These minimum and maximum values are selected based on domain knowledge along with trial and testing in the simulator and are mentioned in Table~\ref{tab:simulation_params}. The optimization problem in \eqref{eq:moop_formul} is non-convex, and therefore, we resort to meta-heuristic algorithms, Simulated Annealing, and Genetic Algorithm, for finding the optimal solution \cite{SA_temp_survey, R07_19_WNDDLMAI_TCOM}. 

\par Simulated annealing (SA) involves several parameters that specify its annealing schedules, such as the initial temperature, cooling schedule, number of iterations, and stopping criteria. The values of these parameters should be chosen such that the objective function is sampled across the entire solution space. Following the discussion in \cite{SA_temp_survey}, we select an adaptive temperature schedule that adjusts the rate of cooling based on the previous runs i.e. $T_{k+1} = \frac{T_k}{1+\left(\ln\left(1+\delta\right)3\sigma T_{k}\right)}$ where, $\sigma$ is the standard deviation of the objective function and $\delta$ is a small real number. Genetic algorithm (GA), on the contrary, relies on bio-evolutionary operations like mutation, crossover, and natural selection to direct the random search into a better solution space and eventually achieve the best solution~\cite{R07_19_WNDDLMAI_TCOM}. 
\subsection{Synthetic Data Generation}\label{sec:uc_synth_simul}
\par To achieve data-driven optimization, the initial step is to train an ML model to learn the relationship between the COPs and network KPIs using data gathered from the network. Subsequently, this ML model is utilized in conjunction with the OE to determine the optimal COP values. To obtain the necessary data, we generate synthetic COP-KPI data using UC-SyntheticNET, a module of the 3GPP compliant system level simulator to model user-centric networks~\cite{R15_20_Syntht_Access}. The simulator considers spatial correlations, as well as mobility, to generate realistic COP-KPI data. We enhance the simulator module by incorporating errors in the positions of UEs and DBSs and simulating their resulting effects on relevant KPIs. This enables the simulator to generate both ideal and erroneous databases.

\par Since ideal data is unavailable at runtime in real networks, it cannot be used for model training in the proposed DD-OEC framework. Instead, the ideal data is considered as historical data, which was obtained in the past through measurement data collection campaigns, such as drive tests. The DD-OEC framework incorporates this data for residual computation to enhance optimization performance. Furthermore, it is also used as a benchmark for performance comparison.
\begin{table}[htb!]
	\setlength{\tabcolsep}{3pt}
	\renewcommand{\arraystretch}{0.99}
	\centering
	\caption{AutoML-based Model Fitting Module Performance: Regression models; GB: Gradient Boosting, LG: Light GBM, RF: Random Forest, CB: Catboost, AB: Adaboost.}
	\label{tab:Auto_ML_Regr}
	\begin{tabular}{|c|ccc|ccc|}
		\hline
		\multicolumn{1}{|l|}{} & \multicolumn{3}{c|}{\textbf{AutoML Model Name}}           & \multicolumn{3}{c|}{\textbf{RMSE Performance}}                              \\ \hline
		\textbf{\begin{tabular}[c]{@{}c@{}}Case\\ Flag\end{tabular}} &
		\multicolumn{1}{c|}{\textbf{\begin{tabular}[c]{@{}c@{}}1st \\ Model\end{tabular}}} &
		\multicolumn{1}{c|}{\textbf{\begin{tabular}[c]{@{}c@{}}2nd \\ Model\end{tabular}}} &
		\textbf{\begin{tabular}[c]{@{}c@{}}3rd \\ Model\end{tabular}} &
		\multicolumn{1}{c|}{\textbf{\begin{tabular}[c]{@{}c@{}}1st\\ Model\end{tabular}}} &
		\multicolumn{1}{c|}{\textbf{\begin{tabular}[c]{@{}c@{}}2nd \\ Model\end{tabular}}} &
		\textbf{\begin{tabular}[c]{@{}c@{}}3rd \\ Model\end{tabular}} \\ \hline
		ASE Model-E                 & \multicolumn{1}{c|}{GBR} & \multicolumn{1}{c|}{LGR} & RFR & \multicolumn{1}{c|}{3.71E-5} & \multicolumn{1}{c|}{3.78E-5} & 3.99E-5 \\ \hline
		EE Model-E                 & \multicolumn{1}{c|}{GBR} & \multicolumn{1}{c|}{LGR} & CBR & \multicolumn{1}{c|}{2.02E-5} & \multicolumn{1}{c|}{2.06E-5} & 2.14E-5 \\ \hline
		\cellcolor[HTML]{FFFFFF}ASE Model-R                 & \multicolumn{1}{c|}{ABR} & \multicolumn{1}{c|}{LGR} & CBR & \multicolumn{1}{c|}{1.02E-5} & \multicolumn{1}{c|}{2.17E-5} & 2.34E-5 \\ \hline
		\cellcolor[HTML]{FFFFFF}EE Model-R                 & \multicolumn{1}{c|}{LGR} & \multicolumn{1}{c|}{ABR} & CBR & \multicolumn{1}{c|}{2.17E-5} & \multicolumn{1}{c|}{2.46E-5} & 2.12E-5 \\ \hline
	\end{tabular}
 \vspace{-0.4cm}
\end{table}
\subsection{DD-OEC Framework}\label{sec:AML_mod_fitting}
The block diagram of the proposed ML-aided DD-OEC optimization framework for data-driven optimization is shown in Fig.~\ref{fig:system-model-framework}. The selected optimization algorithm and data generation methodology in the DD-OEC framework have been previously discussed. Next, we discuss the remaining components, their interconnections, and their reasoning. The DD-OEC framework trains two ML models. The first ML model called \emph{Model-E}, is trained on the erroneous data available at runtime in networks to learn the COP-KPI relation. The COP-KPI relation learned by this model is inaccurate and sub-optimal due to positioning errors. To address this issue, a second ML model, named \emph{Model-R}, is trained to learn the residual error for each input COP. The residual signal is the difference in KPI values for the erroneous and ideal historical data. The output of both models is combined during optimization to compensate for the positioning error and output the COPs, which maximize the desired network KPIs. The residual learning approach is particularly advantageous when dealing with complex relationships between input and output data. Rather than attempting to directly predict the complex, non-linear output, residual learning focuses on learning the residual, which has a limited variation. Training on data with low variance makes the training process more efficient and effective. Furthermore, the impact of noise or errors in the data can be mitigated, which is needed in our scenario \cite{BibResidualLearning}.
\par Since, the main focus of this work is on optimization and error compensation, and not on the ML model fitting, we resort to the off-the-shelf AutoML library, PyCaret, which provides robust regression models for the given data. The ML model fitting problem quite similar to a typical regression problem with three features (DBS density, Szone radius, and transmit power) and two target variables (ASE and EE), and we dedicate a separate model for each target variable. All the standard pre and post-processing steps, like the feature scaling, K-fold cross-validation, and hyper-parameters optimization, are carried out within PyCaret. In addition, PyCaret can identify the best-performing models among the diverse models that are simultaneously trained. The best-performing models, both \textit{Model-E} and \textit{Model-R}, in terms of root mean square error (RMSE) are given in Table~\ref{tab:Auto_ML_Regr}. The results in Table~\ref{tab:Auto_ML_Regr} show that the family of boosting models with their typical hyperparameter values predefined in PyCaret is best suited for the problem under consideration. Specifically, the gradient boosting regressor (GBR) remains the best model with the minimum RMSE values. Moreover, compared to the neural networks these models are faster and requires relatively less data for training. The final key component of the proposed framework is the OE, which employs two widely utilized heuristic optimization algorithms, GA and SA. In the proposed framework, \textit{Model-R} output is added with \textit{Model-E} output within the fitness evaluation functions, of both SA and GA algorithms, for obtaining the COPs which maximize \eqref{eq:obj_func_Eq}.
\begin{table}[h!]
	\setlength{\tabcolsep}{1pt}
	\caption{{Parameters for database generations and simulations.}}
	\centering
	\label{tab:simulation_params}
	\begin{tabular}{ll}
		\toprule
		\textbf{Parameter   Name} 		& \textbf{Value} \\ \midrule
		DBS Density ($\lambda_{dbs}$) 	& 0.0005--0.0125            \\ 
		Transmit Power ($P_{tx}$)       & 15--30   dBm             \\ 
		Szone Radius   ($R_{sz}$)       & 10--50   m               \\ 
		UEs and DBS deployment    		& Poisson Point Process    \\ 
		Positioning Error Distribution  & Uniform    			   \\ 
		Shadowing Standard Deviation                 		& 4    \\ 
		User Density($\lambda_{ue}$)    & 0.0005                   \\ 
		Network Area          		    & 1km   sq                 \\ 
		Bin Size              		    & 10                       \\ 
		Simulation Size      		    & 100 cycles       		   \\ 
		Error Radii          		    & 15   m                   \\ 
		Population Size      		    & 24     			       \\ 
		Initial Temperature, $\delta$, $\sigma$         & 250, 0.0001, 0.01               \\ \bottomrule
	\end{tabular}
 \vspace{-0.5cm}
\end{table}
\begin{figure*}[htb!]
	\centering
	\begin{subfigure}[h]{0.345\textwidth}
		\centering
		\includegraphics[width=7cm, height=5cm]{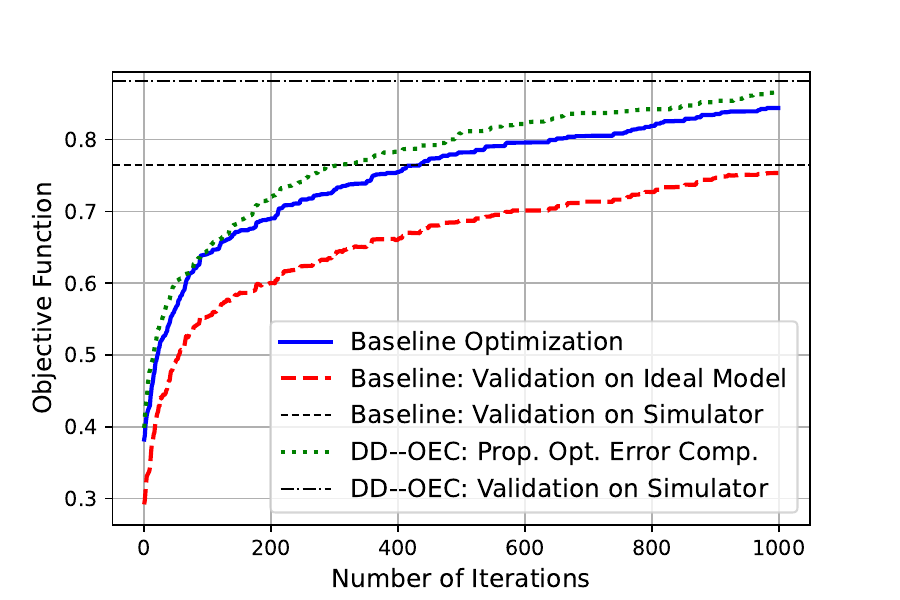}
		\caption{Convergence performance comparison for SA.}
		\label{fig:conver_sa}
	\end{subfigure}
	\begin{subfigure}[h]{0.345\textwidth}
		\centering
		\includegraphics[width=7cm, height=5cm]{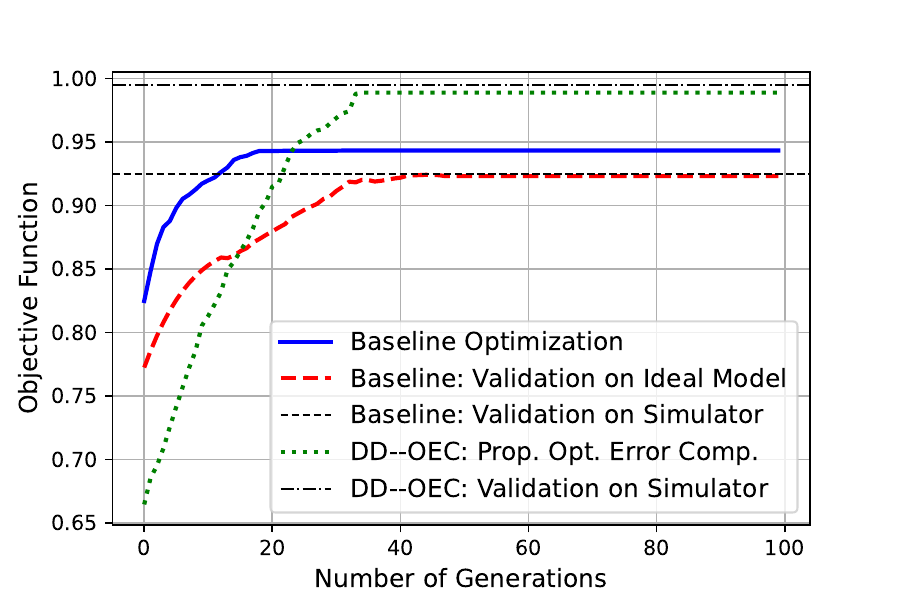}
		\caption{Convergence performance comparison for GA.}
		\label{fig:conver_ga}
	\end{subfigure}
	\begin{subfigure}[h]{0.29\textwidth}
		\centering
		\includegraphics[width=5cm, height=5cm]{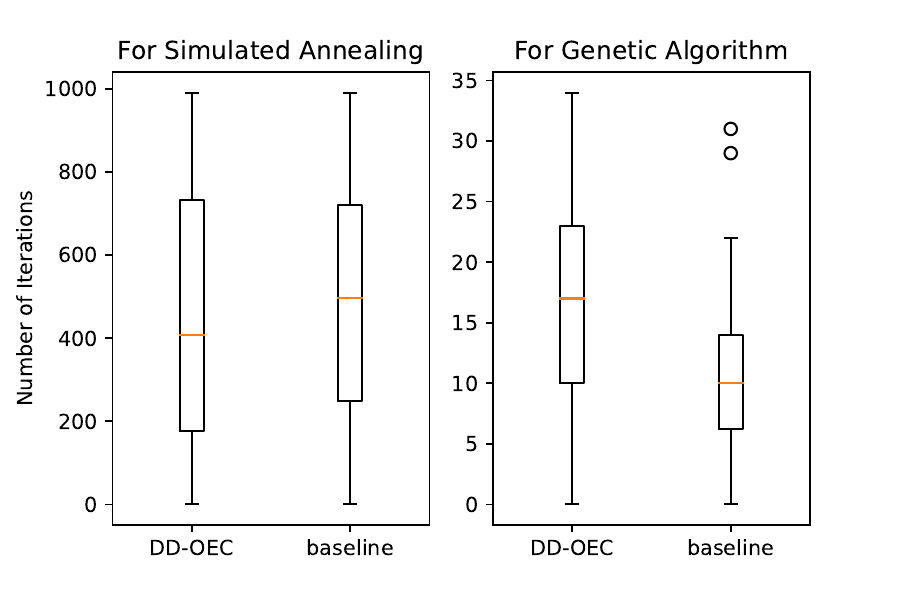}
		\caption{Required iterations for convergence.}
		\label{fig:iterat_sg}
	\end{subfigure}
	\caption{{Performance of the DD-OEC framework with the optimization techniques of SA and GA for $\alpha_{se}=0.5$.}}
	\label{fig:converg_iterat}
 \vspace{-0.5cm}
\end{figure*}
\section{Numerical Simulations \& Performance Evaluation}\label{sec:perform-eval}
\par The performance of the proposed DD-OEC framework is analyzed through numerical simulations. To illustrate the benefit of learning the residual and performing error compensation, the performance of the DD-OEC framework is compared with a \emph{baseline} optimization framework, which does not process the residual and optimizes based on the model solely trained on erroneous data. The performance is evaluated in terms of the achieved objective function values, KPI values, and the corresponding COP combinations. In addition, the number of iterations for convergence is also evaluated. We simulate the user-centric network over a geographic area of 1 km by 1 km. The values of the relevant parameters for network simulation and data generation are given in Table~\ref{tab:simulation_params}. {\color{black} For data generation, $10$ equally spaced values are taken within the ranges specified in Table~\ref{tab:simulation_params}, for each COP parameter. Moreover, as there are three COP parameters, the length of both erroneous and ideal databases becomes $10^3$. Finally, the parameters, $\delta$ and $\sigma$, describe the initial temperature and rate of cooling for simulated annealing.}

\par The convergence performance of SA and GA for both DD-OEC and baseline schemes is shown in Fig. \ref{fig:conver_sa} and Fig. \ref{fig:conver_ga}, respectively. These performance curves are obtained after averaging the converged objective function values from $1000$ Monte-Carlo trials. The COPs obtained after algorithm convergence are validated on the simulator by giving the converged COPs as input to the simulator and using the resulting KPIs to evaluate the objective function in \eqref{eq:moop_formul}. In addition to this, the converged COPs are validated using an ML-based model of the simulator trained using the ideal (without error) COP-KPI combinations. These validated values are plotted as horizontal lines and represent the maximum values after convergence. Moreover, the COPs obtained through the baseline scheme are also validated in each iteration using the ideal model. 
\begin{table*}[t!]
	\centering
        \caption{{Performance comparison of the DD-OEC framework with the baseline scheme for different weights combinations.}}
        \label{tab:optm_sa_ga_weights}
	\resizebox{\linewidth}{!}{
		\begin{tabular}{@{}ccccccccc@{}}
			\toprule
			\multirow{2}{*}{\textbf{\begin{tabular}[c]{@{}c@{}}Opt. \\ Algo. \\ Name\end{tabular}}} &
			\multirow{2}{*}{\textbf{\begin{tabular}[c]{@{}c@{}}Weight\\ Case \\ ($\alpha_{se}$, $1-\alpha_{se}$)\end{tabular}}} &
			\multicolumn{2}{c}{\textbf{Baseline Optimization}} &
			\multirow{2}{*}{\textbf{\begin{tabular}[c]{@{}c@{}}COPs \\ Validation on \\ Simulator\end{tabular}}} &
			\multicolumn{2}{c}{\textbf{DD-OEC Optimization}} &
			\multirow{2}{*}{\textbf{\begin{tabular}[c]{@{}c@{}}COPs \\ Validation on \\ Simulator\end{tabular}}} &
			\multirow{2}{*}{\textbf{\begin{tabular}[c]{@{}c@{}}Exec. \\ Time\\  (s)\end{tabular}}} \\ \cmidrule(lr){3-4} \cmidrule(lr){6-7}
			&
			&
			\multicolumn{1}{c|}{\textbf{\begin{tabular}[c]{@{}c@{}}Objective\\ Function\end{tabular}}} &
			\shortstack{\textbf{Opt. COP Comb.} \\ \newline $[\lambda_{dbs}, R_{sz}, P_{tx} ]$} &
			&
			\multicolumn{1}{c|}{\textbf{\begin{tabular}[c]{@{}c@{}}Objective\\  Function\end{tabular}}} &
			\shortstack{\textbf{Opt. COP Comb.} \\ \newline $[\lambda_{dbs}, R_{sz}, P_{tx} ]$} &
			&
			\\ \midrule
			\multirow{3}{*}{\textbf{SA}} &
			(0.25, 0.75) &
			\multicolumn{1}{c}{0.82=(0.21+0.61)} &
			{[}5.17E-4, 13.33,   24.41{]} &
			0.77=(0.22+0.55) &
			\multicolumn{1}{c}{0.85=(0.23+0.62)} &
			{[}5.79E-4, 10.95,   18.34{]} &
			0.84=(0.23+0.61) &
			433 \\ \cmidrule(l){2-9} 
			&
			(0.50, 0.50) &
			\multicolumn{1}{c}{0.86=(0.45+0.41)} &
			{[}5.74E-4, 11.62,   17.64{]} &
			0.78=(0.44+0.34) &
			\multicolumn{1}{c}{0.87=(0.45+0.42)} &
			{[}6.24E-4, 10.05,   20.64{]} &
			0.87=(0.45+0.42) &
			406 \\ \cmidrule(l){2-9} 
			&
			(0.75, 0.25) &
			\multicolumn{1}{c}{0.84=(0.61+0.23)} &
			{[}5.16E-4, 12.94,   23.31{]} &
			0.79=(0.61+0.18) &
			\multicolumn{1}{c}{0.92=(0.72+0.20)} &
			{[}5.68E-4, 10.70,   16.35{]} &
			0.91=(0.71+0.20) &
			443 \\ \midrule
			\multirow{3}{*}{\textbf{GA}} &
			(0.25, 0.75) &
			\multicolumn{1}{c}{0.88=(0.19+0.69)} &
			{[}5.05E-4, 12.26,   17.23{]} &
			0.72=(0.16+0.56) &
			\multicolumn{1}{c}{0.89=(0.18+0.71)} &
			{[}5.89E-4, 11.23,   20.36{]} &
			0.89=(0.18+0.71) &
			313 \\ \cmidrule(l){2-9} 
			&
			(0.50, 0.50) &
			\multicolumn{1}{c}{0.94=(0.46+0.48)} &
			{[}5.50E-4, 11.63,   15.25{]} &
			0.92=(0.48+0.44) &
			\multicolumn{1}{c}{0.98=(0.49+0.49)} &
			{[}5.35E-4, 10.16,   15.00{]} &
			0.96=(0.49+0.47) &
			306 \\ \cmidrule(l){2-9} 
			&
			(0.75, 0.25) &
			\multicolumn{1}{c}{0.88=(0.69+0.19)} &
			{[}5.20E-4, 13.24,   23.31{]} &
			0.81=(0.62+0.19) &
			\multicolumn{1}{c}{0.87=(0.72+0.15)} &
			{[}5.75E-4, 11.07,   18.85{]} &
			0.86=(0.72+0.14) &
			316 \\ \bottomrule
		\end{tabular}
	}
 \vspace{-0.6cm}
\end{table*}
\par It can be observed from Fig. \ref{fig:conver_sa} and Fig. \ref{fig:conver_ga} that even though the baseline scheme improves the objective function value with erroneous data (see blue curve) when these COPs are validated using the ideal model (red curve) or the simulator, the objective function value is much lower. This indicates that with erroneous data, the COPs obtained after optimization are highly sub-optimal for the actual network. On the contrary, the COPs obtained via the DD-OEC scheme achieve the maximum value of the objective function when validated using the simulator. These results reveal that the mechanism of DD-OEC is necessary for the attainment of a close-to-optimal COP combination, resulting in a better network operating point. 

\par Moreover, it can be observed that the convergence of GA differs from SA convergence and is able to converge earlier. SA shows a rising trend because of a high degree of variability of its convergence path in each run. Contrary to this, the convergence path of GA is relatively less varying over multiple runs. Furthermore, as GA is a population-based algorithm, which remains fixed in each run, its initialization is not as random as that of SA, and it reaches the optimum solution in relatively less time. In addition, the number of generations/iterations required to reach the optimum solution varies in different runs of the same algorithm. Hence, to provide more insights, we plot these results using box plots in Fig. \ref{fig:iterat_sg}. An important observation to note here is that the number of required iterations in the DD-OEC scheme is slightly higher than the baseline scheme because the DD-OEC scheme needs to combine the effect of both the \textit{Model-E} and \textit{Model-R}; hence it takes more time to converge. 

\par The simulation results with different values of $\alpha_{se}$ and $1-\alpha_{se}$ are summarized and compared in Table~\ref{tab:optm_sa_ga_weights}. {\color{black}For both the baseline and proposed DD-OEC schemes, the achieved objective function value along with the corresponding EE and ASE values and the respective COP values, indicated by the triplet $[\lambda_{dbs}$, $R_{sz}$, $P_{tx}]$, are also shown in Table~\ref{tab:optm_sa_ga_weights}.} Again, it can be noted that for both SA and GA, the objective function value achieved for the baseline scheme is lower compared to the objective function value for the DD-OEC scheme. When the baseline COPs are validated using the simulator, the objective function value is lowered further, highlighting the sub-optimality of the solution. On the contrary, when the optimized COPs from the DD-OEC framework are validated on the simulator, the utility degradation is minimal. Compared to the baseline scheme, it can be noted that with the SA scheme, a gain up to $15\%$ is observed, and with the GA scheme a gain of up to $23\%$ is achieved. This highlights the need for the DD-OEC scheme and validates its effectiveness to counter the impact of errors. 

\par Furthermore, it can be noted that the optimized COPs for different cases converge to lower values of $\lambda_{dbs}$ and $R_{sz}$ and medium values for the transmit power. This is because, for the lower Szone sizes, relatively more UEs are scheduled, which increases the ASE. A lower Szone radius results in an increase in interference, however, this consequence is avoided by reducing the transmit power. The lower value of the DBS density is due to the limited UE density and this lower DBS density is sufficient to serve the limited UE density. 
Comparing the different values of $\alpha_{se}$, it can be noted that the maximum objective function value is obtained when both ASE and EE are given equal weight as both components are normalized. Finally, comparing GA and SA reveals that the former, with its population-based solution-finding approach, can efficiently reach convergence, and the converged objective function values are also higher. Hence, for this problem, GA is the better choice.
\section{Conclusion}\label{sec:conclusion}
The errors in UE and DBS positions impact the performance of user-centric networks. In particular, the performance of the data-driven optimization solutions becomes sub-optimal when utilizing the data generated from the UEs with erroneous positions. This paper proposes a data-driven framework to minimize the adverse impact of UE and DBS positioning errors on the performance of the user-centric network. The framework relies on an error residual model trained on historical data, which can be collected from a drive test campaign. Data-driven models from AutoML are trained on the current erroneous data to learn the impact of parameters on spectral and energy efficiency. A multi-objective optimization problem of joint maximization of spectral and energy efficiency was formulated. We then utilize the current erroneous model and the historical residual model in the OE to predict the optimum COPs. Results indicate that the proposed DD-OEC method produces better convergence compared to the baseline optimization method.\vspace{-0.25cm}
\section{Acknowledgement}
This work is supported by the National Science Foundation under Grant Numbers 1923669, 2323300, 1923295.

\end{document}